# Eliminating edge electronic and phonon states of phosphorene nanoribbon by unique edge reconstruction

Shi-Qi LI[1+], Xiangjun Liu[2+], Xujun Wang[2], Hongsheng Liu[1], Gang Zhang[3]*, Jijun Zhao[1], Junfeng Gao[1,3]*

1. Key Laboratory of Materials Modification by Laser, Ion and Electron Beams (Dalian University of Technology), Ministry of Education, Dalian 116024, China
2. State Key Laboratory for Modification of Chemical Fibers and Polymer Materials, Institute of Micro-/Nano Electromechanical System, College of Mechanical Engineering, Donghua University, Shanghai, China
3. Institute of High-Performance Computing, A*STAR, Singapore

**Abstract.** Edge termination plays a vital role in determining the properties of 2D materials. By performing compelling *ab initio* simulations, a lowest-energy **U**-edge [ZZ(U)] reconstruction is revealed in the bilayer phosphorene. Such reconstruction reduces 60% edge energy compared with the pristine one and occurs almost without energy barrier, implying it should be the dominating edge in reality. The electronic band structure of phosphorene nanoribbon with such reconstruction resembles that of intrinsic 2D layer, exhibiting nearly edgeless band characteristics. Although ZZ(U) changes the topology of phosphorene nanoribbon (PNR), simulated TEM, STEM and STM images indicates it is very hard to be identified. One possible identify method is IR/Raman analyses because ZZ(U) edge alters vibrational modes dramatically. Beyond, it also increases the thermal conductivity of PNR 1.4 and 2.3 times than the pristine and Klein edges.

**Key words**：Black phosphorene, edge reconstruction, band structure, thermal transport, electronic properties

Phosphorene, the mono- or few-layer of black phosphorus, has gained tremendous research attention [1-3] since it was produced by mechanical [1, 2], liquid [4, 5], and electrochemical exfoliation method [6]. Phosphorene is an ideal two-dimensional (2D) material for ultrafast microelectronics because it possesses both high on/off ratio [3] and extremely high carrier mobility [3, 7]. Phosphorene is also proposed to be good thermal transport controlling material [8, 9] due to its high anisotropy in thermal and electronic transport [10]. Besides, phosphorene can be an excellent active anode material for high-rate high-capacity lithium battery. [11] In real production, fabrication and application, phosphorene device should be in nanometer size, where the edge could influence the electronic [12, 13], magnetic [14, 15], and topological [16, 17] properties greatly. For example, graphene zigzag edge brings antiferromagnetic properties [14, 18-20], while its armchair edged ribbons exhibit a magic dependence on width [20].

Generally speaking, edge breaks the perfect bonds situation and results in high edge energy. To passivate the dangling bonds, reconstruction is likely to occur at the edge of 2D materials [21-23]. The edge reconstruction has pronounced impact on the electronic/magnetic properties, especially for those narrow nanoribbons [24, 25]. For example, the ZZ(57) and AC(677) reconstructions at graphene edge were firstly proposed by theoretical study [21] and then identified by experimental microscopy [26, 27]. Similarly, there are (2×1), (3×1) [23] and DT edge reconstructions [28] for transition metal dichalcogenides (TMDCs), resulting in distinctively different electronic properties and exotic magnetic orderings.

For phosphorene, zigzag (ZZ) direction is the dominating edge according to recent experiments [29]. Most of previous studies focused on the edge of monolayer phosphorene [30-34]. However, usually the phosphorene samples synthesized in experiments are multilayered. Different from graphene [21, 26, 27, 35] and $MoS_2$ [28], there is still a big gap of knowledge about the edge structure and influence for bilayer (2L) phosphorene.

In this letter, we disclose a novel **U**-edge (denoted as [**ZZ(U)**] edge in the following) for bilayer phosphorene nanoribbon (PNR). ZZ(U) shows nearly edgeless states in band gap demonstrated by systematic *ab initio* exploration. Besides, ZZ(U) greatly reduces the edge energy by 60% and the reconstruction is nearly barrierless. All typical vibrational mode shifts and TEM, STEM, STM images of **U**-edge are simulated for further experimental identification. Such a **U-**edge makes Fermion in bilayer PNR is insensitive to the edge, originated from eliminated edge state in band structure. More interesting, the ZZ(U) reconstruction can significantly increase thermal conductivity of bilayer PNR, nearly two-fold enhancement compared with the Klein edge. The underlying mechanism is discussed based on local heat flux distribution.

First, we explore the bilayer edge (Fig. 1 d-h) of phosphorene by composing of the three mostly studied monolayer edge structures, as shown in Fig.1 (a)-(c), i.e. pristine ZZ edge [ZZ(Pristine)], Klein (Cliff) edge [ZZ(Klein)], and tube-terminated edge [ZZ(Tube)]. As mentioned before, ZZ(Tube) is the most stable one for monolayer [32]. However, the height of ZZ(Tube) (3.38 Å) is remarkably higher than that of central section of phosphorene edges (2.11 Å) (Fig. S1a). Therefore, bilayer PNR with ZZ(Tube) edge is unstable because the large interlayer gap (6.35 Å) will decouple bilayer structure to two individual sheets. The structure in Fig. 1g formed by half ZZ(Tube) and half [ZZ(Pristine)] is also less stable than the pure [ZZ(Pristine)] (Fig. 1e) and pure ZZ(Klein) (Fig. 1f) edge. Therefore, ZZ(Tube) is favored for monolayer phosphorene, but is not for bilayer phosphorene.

Intriguingly, when the upper and bottom layers are staggered in half zigzag period, the separated ZZ edges can spontaneously connect to each other, forming a self-passivating folding edge (Fig. 1d). In this letter, we denote this folding edge as **U**-edge [**ZZ(U)**], as shown in Fig.1h. Notably, the ZZ(U) is significantly more stable than other types of edge. To compare the stability of these four bilayer edges, their edge energy per length $\gamma_{2L}$ are calculated as $\gamma_{2L} = \frac{1}{4L}(E_{PNR} - N_P E_P)$, where $E_{PNR}$ is the total energy of a bilayer PNR, $N_P$ is the number of P atoms in the bilayer PNR, and $E_P$ is the energy of a P atom in the perfect bilayer phosphorene. *L* represents the length of edges and the factor 4 is the number of edges in bilayer pristine PNR.

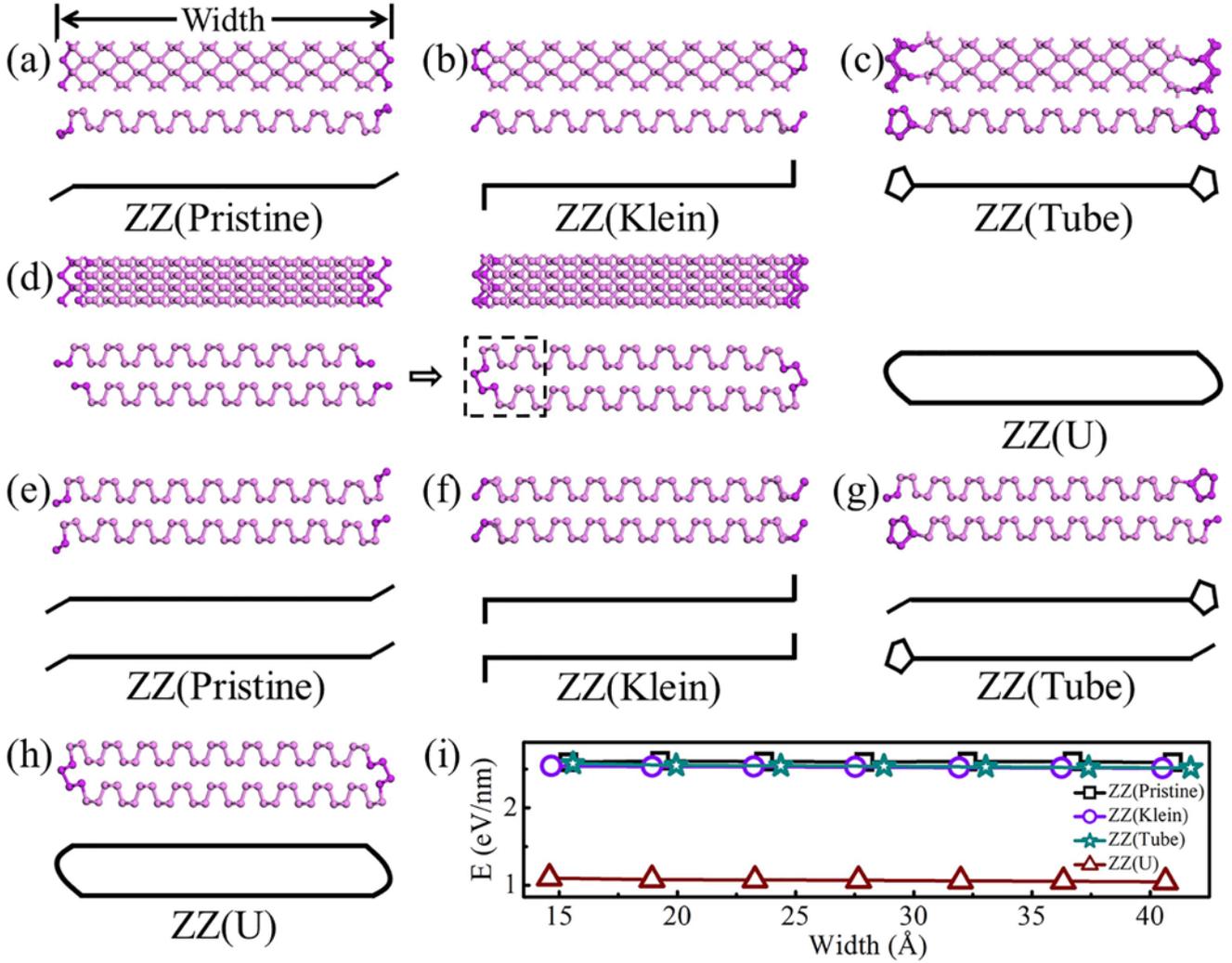

FIG. 1. Schematic of edge reconstruction patterns of bilayer phosphorene nanoribbons. Three possible edge structures and corresponding schematics of monolayer phosphorene including ZZ(Pristine) (a), ZZ(Klein) (b), ZZ(Tube) (c). The barrierless formation process of U edge (d). Four possible edge reconstructions for bilayer phosphorene (e-h) and their related edge energies (i). The definition of width is shown in (a).

Clearly, ZZ(U) is the lowest-energy edge among all explored bilayer edges for the PNRs with width from 14.58 to 40.64 Å (Fig. 1i). ZZ(U) reduces the edge energy by over 59.7% compared with ZZ(Pristine). Moreover, the reconstruction pattern of bilayer PNRs with four ZZ(Tube) edge is also considered. Its edge energy is also significantly higher than ZZ(U) as expected (Fig. S1b). Considering the lowest-energy and barrierless kinetic, ZZ(U) could be the ground state edge of bilayer PNR. Furthermore, we also perform *ab initio* molecular dynamics (AIMD) simulation for ZZ(U) edge to verify its thermal stability. During the AIMD simulation with NVT ensemble at 300 K (Fig. S2 and Movie-S12), the total energy oscillates with amplitude of less than 31 meV per atom for ZZ(U) edge. After 5 ps simulations, the proposed geometric structures of 2L PNRs are evidently retained (Movie-S12). All these facts suggest PNRs terminated with U-edge are thermodynamically stable at room temperature.

This unprecedented **U**-edge may be a unique edge reconstruction of 2D structure. Black phosphorene has distinct puckering lattice from graphene and TMDCs. Upon transforming into **U**-edge, the unsaturation edge atoms are all passivated since all P atoms form perfect three bonds with a lone pair of electrons. Only slight penalty comes from the change of orientation for a lone pair, while the interlayer distance has neglectable change. Therefore, U-edge greatly reduces the edge energy.

Nowadays, black phosphorene may be produced by exfoliation, the U-edge closes the border of phosphorene, impeding the ion intercalation. Therefore, the U-edge will limit the production of monolayer phosphorene. This may provide explanation why monolayer black phosphorene is so rare, and most phosphorene are few-layers in experiments [36, 37]. Besides, for epitaxial growth of phosphorene, the presence of large amount of such U-edge will affect how the atoms dock and consequently, the equilibrium morphologies of phosphorene islands [38-41]. Besides, the interaction between the nanoflake terminated with U-edge and substrate will become a critical factor for selecting a suitable substrate for growth of phosphorene [42].

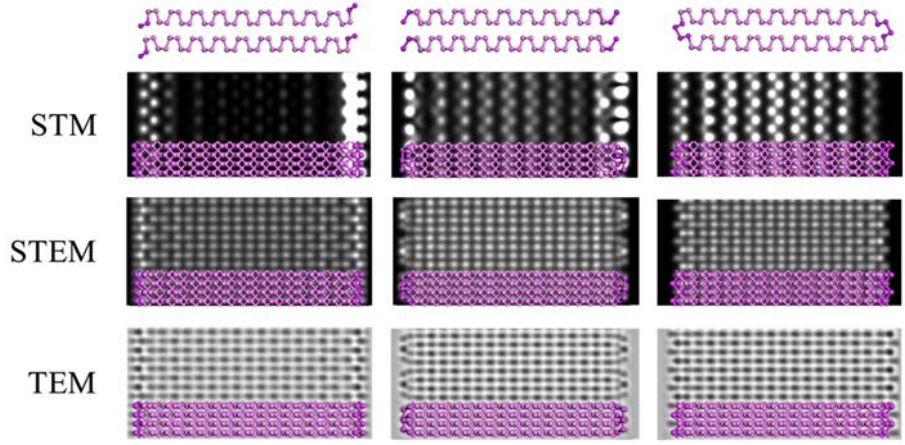

FIG. 2. Simulated STM images at –0.4 V bias (upper panel), STEM images at 200 kV (middle panel), and TEM images at 200 kV (lower panel) for ZZ(Pristine), ZZ(Klein), and ZZ(U), respectively.

Generally speaking, edge greatly influences the electronic properties of nanoribbons, which are commonly observed and well-studied in graphene [21] and TMDCs [23]. Especially, strong edge states are common in ZZ direction because it is hard to passivate [21]. Therefore, scanning tunneling microscopy (STM) is popularly used to observe the local edge structures. However, nearly all the previous experimental reports cannot observe the pronounced edge states on PNR, distinctly different from the other 2D materials [43-45].

We simulate the STM, scanning transmission electron microscope (STEM), transmission electron microscope (TEM) images of ZZ(Pristine), ZZ(Klein) and ZZ(U), respectively (see Fig 2). Obviously, both ZZ(Pristine) and ZZ(Klein) present clear edge state in simulated STM images. However, there is totally no edge state for ZZ(U) edge in the studied three microscope images. This explains the fact that the edge state of PNR is hardly observed experimentally. This phenomenon inspires deep reconsideration about the real edge structures of all the previous experimental observation.

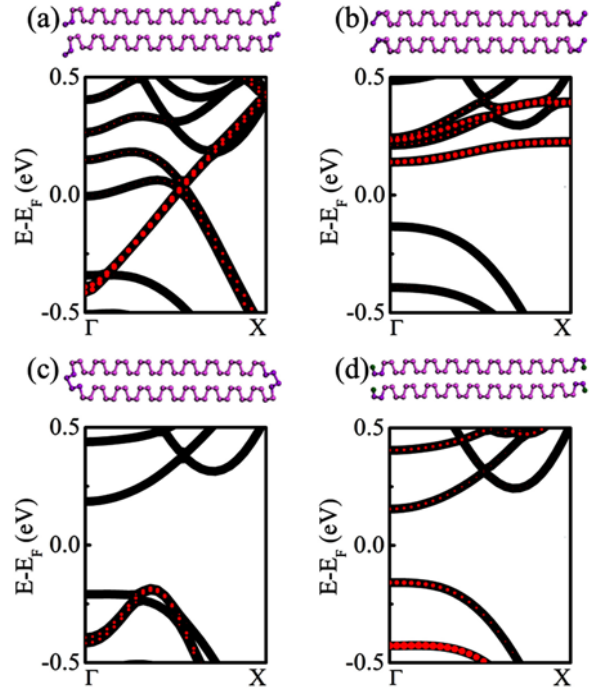

FIG. 3. The electronic band structures with highlighted edge state as red dots for ZZ(Pristine) (a), ZZ(Klein) (b), ZZ(U) (c), and the hydrogenated ZZ(H) (d).

Such a U-edge connects two layers into one unity, changing the topological structure of PNR. Its electronic property is certainly different from the pristine edge. As shown in Fig. 3a and Fig. S3, bilayer PNR with ZZ(Pristine) always exhibits typical metallic behavior. Meanwhile, PNR with Klein edge is a semiconductor with the edge states locate near the conduction band (Fig. 3b and Fig. S4). In contrast, the PNR with ZZ(U) is just like a flat nanotube, which has none edge in the radial direction. Therefore, the carrier of ZZ(U) can be regarded as edgeless, prohibiting edge scattering. We further explored the band structure of the U-edge PNR with different widths (Fig. S5). It is found the U-edge state is like to sink into valence band as the width increasing. When the PNR is up to 40.64 Å (Fig. 3c), the edge state of U-edge nearly vanished in the band gap. At $\Gamma$ point, edge state locates 0.2 eV lower than VBM. Then edge state first emerges and goes down from $\Gamma$ to X, and only very limited proportion is slightly above the top of valence band of bulk phosphorene. For more longer PNR with ZZ(U) edge, the edge state should be hidden in bulk band structure, resembling band characteristic of the perfect bilayer phosphorene (Fig. S6), not sensitive to the width. Hydrogenated ZZ(H) edge is compared calculated in (Fig.3d, Figure S7). Although ZZ(H) is semiconducting edge, the edge charge contributes both to highest valence band and lowest conduction band. The band gap of ZZ(H) PNR is more sensitive to the width than that of ZZ(U).

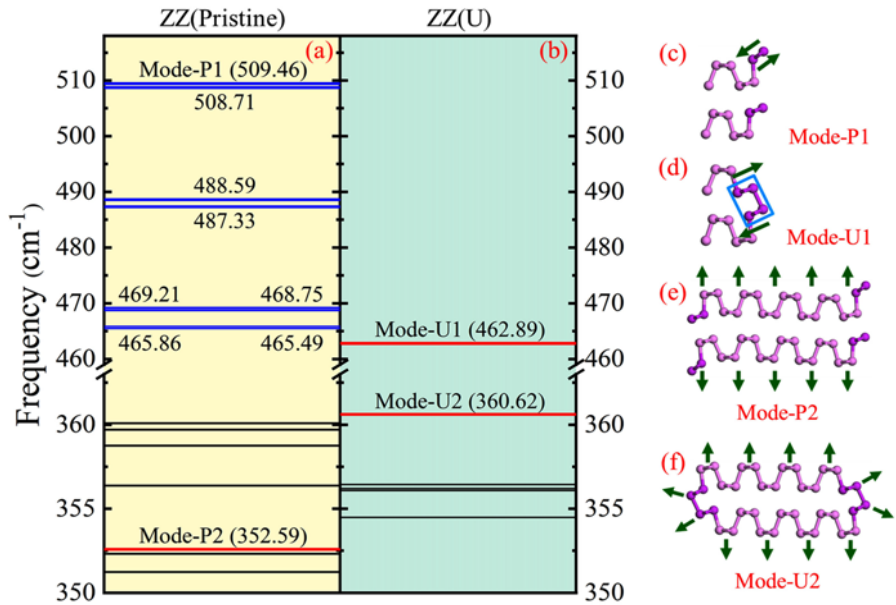

FIG. 4. Vibration frequencies for ZZ(Pristine) (a) and ZZ(U) (b). Typical sole vibration frequencies for ZZ(Pristine) and ZZ(U) are labeled as blue and red lines, respectively. (c)-(i) Some representative vibrating modes denoted in (a)-(b).

As the usually used STM, STEM and TEM technologies are hardly to identify the newly reported ZZ(U), next we provide possible experimental criteria, i.e., infrared (IR) or Raman spectroscopy for structural identification. We compare the vibrational frequency for ZZ(Pristine) and ZZ(U) (Fig. S8) in detail, and find that the difference mainly concentrates within two frequency ranges: 459-518 and 350-363 cm$^{-1}$, as shown in Fig. 4.

At high-frequency region (465, 510) cm$^{-1}$, there are about eight vibration modes (Fig.4a and Movie S13-20) corresponding to stretching and twisting vibration of the edge atoms of separated ZZ(Pristine). The highest vibration frequency locates at 509.46 (Fig.4a, 4c). After reconstruction to form the ZZ(U) edge, separated ZZ(Pristine) changes to bonded ZZ(U) edge, which results in edge stretching modes. The highest vibration frequency for ZZ(U) edge is only 462.89 cm$^{-1}$ (Fig. 4b and Mode-U1 in Fig.4d, Movie S22), while the vibration modes of ZZ(Pristine) disappear. Therefore, measurement of the edge vibration frequency such as IR spectrum is an effective way to reveal the ZZ(U) edge reconstruction.

Beyond, the ZZ(U) edge connects two phosphorene layers together, therefore some relative vibrations between layers vanish, such as interlayer relative vertical vibrating mode (Mode-P2 in Fig. 4e and Movie S21) with frequency of 352.59 cm$^{-1}$ (Fig. 4a). Instead, a breath mode (Fig. 4f, Movie S23) appears with frequency of 360.62 cm$^{-1}$ (Fig 4b), which is expected to exhibit remarkable Raman activity.

Such different vibrational modes can be expected to affect the thermal conductivity of bilayer PNRs, which are comparingly investigated using equilibrium molecular dynamics (MD) simulations. The three bilayer PNRs with the same size of width (40.64 Å) and length (103.68 Å) are considered, and the well-established Stillinger-Weber (SW) potential is used to describe the interactions between P atoms [46, 47]. The thermal conductivities $\kappa$ of bilayer PNRs along the ZZ direction are calculated with the Green-Kubo formula, which was derived from linear response theory and fluctuation-dissipation theorem. In Green-Kubo formula, the thermal conductivity is related to heat flux autocorrelation function [48], $\kappa = \frac{V}{K_B T^2} \int_0^\infty \langle J_i(0) J_i(t) \rangle dt$, where $V$ denotes the system volume, $K_B$ is the Boltzmann constant, $\langle J_i(0) J_i(t) \rangle$ is the average heat flux autocorrelation function, and $J$ represents the heat flux, $J = \frac{1}{V} \left[ \sum_i^N \varepsilon_i v_i + \frac{1}{2} \sum_{ij; i \neq j}^N (F_{ij} \cdot v_i) r_{ij} + \frac{1}{6} \sum_{ijk; i \neq j \neq k}^N (F_{ijk} \cdot v_i)(r_{ij} + r_{ik}) \right]$, where $\varepsilon_i$ and $v_i$ denote the energy and velocity of atom $i$, respectively. $r_{ij}$ is the distance between atom $i$ and atom $j$. $F_{ij}$ and $F_{ijk}$ are the multi-body force vector between atoms $i$, $j$ and i, j, k.

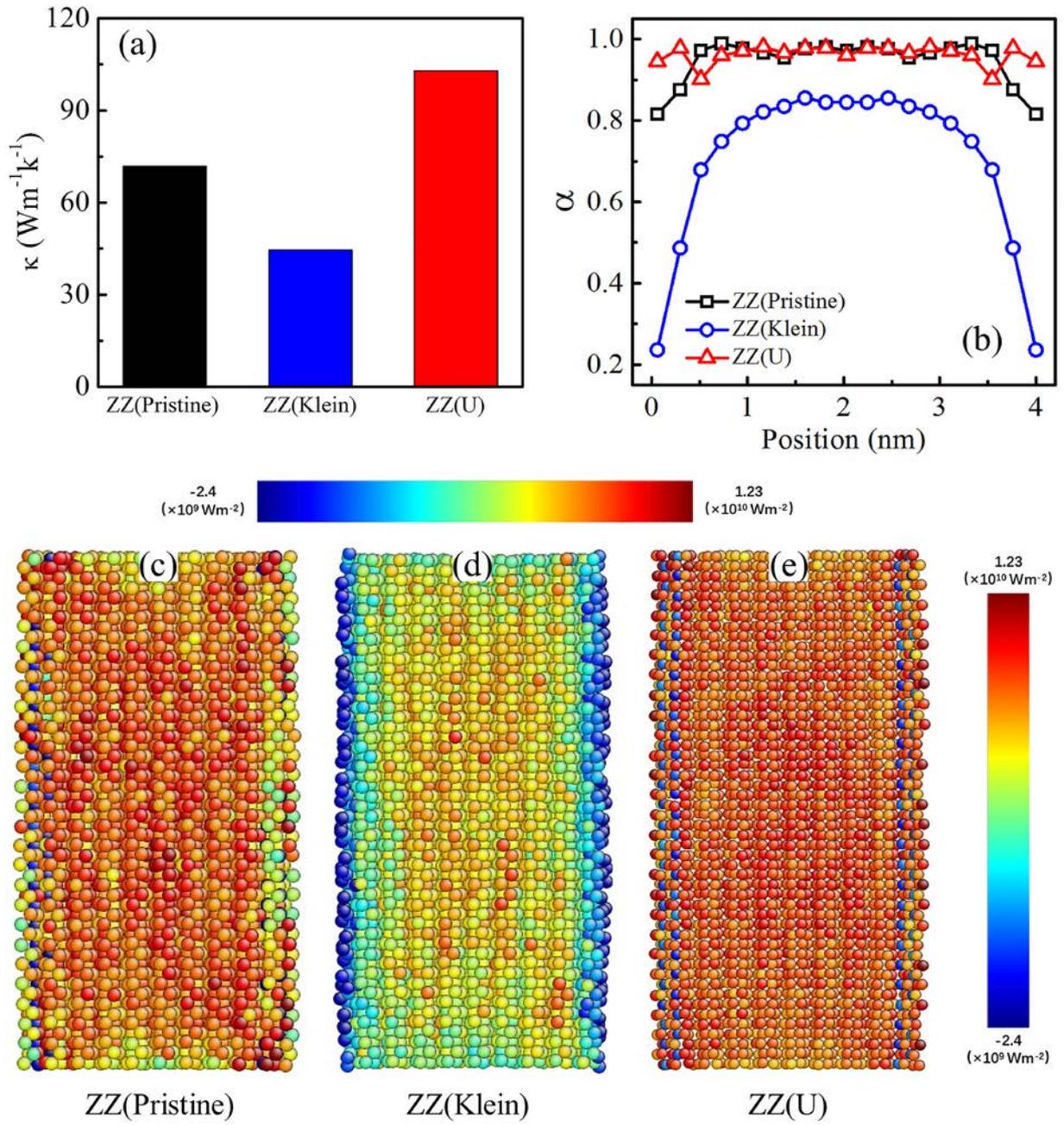

FIG. 5. The thermal conductivity (a) and normalized heat flux density distribution (b) of PNR with ZZ(Pristine), ZZ(Klein) and ZZ(U) edges. The heat flux density distribution of PNR with ZZ(Pristine) (c), ZZ(Klein) (d), and ZZ(U) (e) edges are also shown.

Size effect on the thermal conductivity of nanostructure is a very important issue for the applications of 2D materials. To clarify the size effect on the thermal conductivities of PNRs, corresponding PNRs with larger width (62.28 Å) have been also considered. As shown in Figure S9, the thermal conductivities of PNRs are size dependent, which increase as the width of PNRs increases. But the relative trend in thermal conductivity will not change. Therefore, we will focus on bilayer PNRs with width of 40.64 Å. Here the calculated thermal conductivity of perfect bilayer BP is $\kappa_{2D} = 146.92$ Wm$^{-1}$K$^{-1}$.

ZZ(U) exhibits significantly higher thermal conductivity than the other two, as shown in Fig. 5a. The thermal conductivity of PNR with ZZ(U) is 102.95 Wm$^{-1}$K$^{-1}$, while that of ZZ(Pristine) PNR and ZZ(Klein) PNR is 71.94 Wm$^{-1}$K$^{-1}$ and 44.52 Wm$^{-1}$K$^{-1}$, respectively. The normalized heat current autocorrelation function (HCACF) and thermal conductivity as a function of correlation time of these three edges are also provided in the Figure S10 and Figure S11, respectively. Thermal conductivity of ZZ(U) is at least 1.4 and 2.3 times of the pristine and Klein edges, indicating U-edge dramatically enhances the thermal conductivity of PNRs.

To understand the remarkable increase in thermal conductivity of ZZ(U) edge, we explored the cross-sectional distribution of the heat flux density in the PNRs using non-equilibrium molecular dynamics simulation. The atoms close to the two ends of PNRs were placed into hot and cold Nosé-Hoover reservoirs with temperatures set to be $T_H = 310$ K and $T_C = 290$ K, respectively. In the calculations, the cross-section of the PNRs is uniformly divided into ninteen columns. As shown in Fig. 5c-e, the heat flux is spatially inhomogeneously distributed. To compare quantitatively the cross-section distribution of the local heat flux along the three PNRs, we introduce the inhomogeneity ratio $\alpha$ which is defined as $\alpha = J_{local} / J_{mid}$, where $J_{local}$ is the local heat flux density in the local regime of the considered PNR, and $J_{mid}$ is the local heat flux in the central regime of pristine PNR. The position-dependent inhomogeneity ratio for PNR with pristine edge, Klein edge and U-edge is presented in Fig. 5b. It is clear that for pristine PNR, the local heat flux at the edge is only ~80% of that at the center, and for Klein PNR, the local heat flux at edge is only about 20% of that at the center, revealing a stronger edge effect. The significant reducion in local heat flux at edge is attributed from strong phonon edge scattering, which is a general phenomenon in nanoribbons [48, 49]. Hence, the localized phonons at edge are responsible for the reduction of the local heat flux. However, in the ZZ(U) edge, the local heat flux at edge is close to that at the center, revealing suppressed phonon edge scattering. More interestingly, in ZZ(U) PNR, the lowest heat flux ($\alpha$ ~90%) is at the next-to-outmost regime instead of the outmost region, that is, the region bridging the U-edge and the central part. The local heat flux at the U-edge is about 95% of that at the central regime, which is remarkably higher than that at the pristine edge and Klein edge. At edge of nanoribbons, the bond between under-coordinated edge atoms is shorter and stronger with respect to those at the central region. The change in local bond strength provides perturbation to the local potential, and subsequently traps phonon transport [51]. Clearly, the ZZ(U) edge maintains the bond strength at edge, thus results in slight impact on phonon transport.

It should be noted that in our work, the heat flux formula is based on virial stress tensor, which is the default heat current formula implemented in LAMMPS. Recently, Fan et al. demonstrated that there is more accurate heat flux formula for Stillinger-Weber (SW) potential [52]. Compared with the commonly used heat flux formula, extra terms containing three-body force are included in the newly suggested expression. However, as our concerns in the present work are the relative change in thermal conductivity with different edge configurations and the heat flux distribution across nanoribbon, instead of the absolute values, here we leave the MD calculation with the corrected many-body heat flux formula in future work.

In summary, we disclose the unique reconstruction, ZZ(U) edge, of phosphorene bilayer for the first time. ZZ(U) edge reduces about 60% edge energy than pristine ZZ edge. AIMD simulations verify the thermodynamic stability of **U**-edge at room temperature. ZZ(U) eliminates high vibrations (> 469 cm$^{-1}$) and low relative vibrations between layers, but brings new vibration modes from bonding edge atoms and generates a breath mode. ZZ(U) maintains the bond strength, suppressing phonon edge scattering and resulting in remarkably higher inhomogeneity ratio for heat flux density. ZZ(U) edge enhances the thermal conductivity by at least 1.4 and 2.3 times than the pristine and Klein edges. We also provide simulated STM, STEM and TEM images for experimentalists to recognize these edges. U-edge is formed through the small structural reconstruction of puckering lattice of bilayer black phosphorene. Therefore, 2D materials with similar puckering lattice to black phosphorene may have similar U-edge reconstruction, which certainly deserves further study. Beyond, we indicate ZZ(U) bring nearly edgeless carriers, which may provide new applications features of PNRs in electric, thermal, optoelectric and catalysis.

**Computational Method**

All the ab initio calculations in this work are performed using plane-wave pseudopotential technique as implemented in the Vienna Ab-initio Simulation Package (VASP) [53]. The exchange–correlation interaction was described by generalized gradient approximation (GGA) with the Perdew–Burke–Ernzerh of (PBE) functional [54]. Core electrons were described by the projector-augmented wave (PAW) method [55]. Grimme's DFT-D3 with Becke-Jonson damping approach was used to describe the vdW interaction [56]. A plane-wave basis kinetic energy cutoff was taken as 500 eV and a convergence criterion of $10^{-5}$ eV were adopted in all the calculations. In the present work, we employ a unit cell containing two periods along the edge for studying the edge reconstruction. The criteria for the convergences of force between atoms were set to 0.01 eV/Å. In the *ab initio* molecular dynamics (AIMD) simulations, 1×2×1 supercells are adopted to minimize the constraint of periodic boundary conditions with the time step of 1fs in the moles–volume–temperature (NVT) ensemble, and the temperature is first heated up to 300K for 1ps and then kept at 300 K for 5 ps. The vibration patterns are visualized by Jmol [57].

Molecular dynamics simulations are employed to study the thermal property of PNRs using the LAMMPS package [58]. During the simulations, the atomic motion is calculated from the integration of Newton's equations with velocity Verlet algorithm, and the time step is set as 0.5 fs. First, the system was relaxed under isothermal–isobaric (NPT) ensemble, and then equilibrated under canonical NVT (constant atom number, volume and temperature) ensemble with a temperature at 300 K for 200 ps using Nosé-Hoover thermostat [59]. Next, the system was further relaxed under NVE (constant atom number and volume, and no thermostat) ensemble for additional 200 ps, which is long enough for the system to reach equilibrium state. Then, the thermal conductivities κ of bilayer PNRs are calculated with the Green-Kubo formula [48].

**Acknowledgements**

This work is supported by the National Natural Science Foundation of China (Grant No. 12074053, 91961204, 12004064) and by XinLiaoYingCai Project of Liaoning province, China (XLYC1907163). J. G. thanks the funding support from Jiangsu Key Laboratory for Carbon-Based Functional Materials & Devices, Soochow University. H. L. thanks the start-up funding (DUT20RC(3)026). We thank the Dr C. Jin and F. Yao for meaningful discussion. We also acknowledge Computers supporting from Shanghai Supercomputer Center, DUT supercomputing center, and Tianhe supercomputer of Tianjin center.

+ These authors have the same contribution to this work.
*Corresponding authors. gaojf@dlut.edu.cn; zhangg@ihpc.a-star.edu.sg

Supplemental Material for the paper

# Eliminating edge electronic and phonon states of phosphorene nanoribbon by unique edge reconstruction

by Shi-Qi LI[+], Xiangjun Liu[+], Xujun Wang, Hongsheng Liu, Gang Zhang[*], Jijun Zhao, and Junfeng Gao[*]

S1. Height of ZZ(Tube) and edge energies of five possible edge reconstructions for bilayer phosphorene

S2. Evolution of temperature and total energy in the ab initio MD simulations

S3. Band structure of ZZ(Pristine) with width from 15.39Å to 40.96Å

S4. Band structures of ZZ(Klein) with width from 14.67 Å to 40.52 Å

S5. Band structures of ZZ(U) with width from 14.58 Å to 40.64 Å

S6. Band structure of the perfect 2L phosphorene

S7. Band structures of hydrogenated ZZ(Pristine) with width from 17.27 Å to 43.15 Å

S8. All the vibration frequencies for ZZ(Pristine) and ZZ(U) within (0, 525) cm$^{-1}$

S9. Size effect on the thermal conductivity of PNRs

S10. Heat current autocorrelation function (HCACF) for pristine edge, Klein edge, and U-edge

S11. The thermal conductivity as a function of correlation time for pristine edge, Klein edge, and U-edge

S12. (Movie) Snapshots for the bilayer ZZ(U) during the *ab initio* molecular dynamics simulations under 300K.

S13. (Movie) **Mode-P1**: The ZZ(Pristine) edge related stretching vibration mode with frequency of 509.46 cm$^{-1}$.

S14. (Movie): The ZZ(Pristine) edge related stretching vibration mode with frequency of 508.71 cm$^{-1}$.

S15. (Movie): The ZZ(Pristine) related stretching vibration mode with frequency of 488.59 cm$^{-1}$.

S16. (Movie): The ZZ(Pristine) related stretching vibration mode with frequency of 487.33 cm$^{-1}$.

S17. (Movie): The ZZ(Pristine) related stretching and twisting vibration mode with frequency of 469.21 cm$^{-1}$.

S18. (Movie): The ZZ(Pristine) related stretching and twisting vibration mode with frequency of 468.75 cm$^{-1}$.

S19. (Movie): The ZZ(Pristine) related stretching and twisting vibration mode with frequency of 465.86 cm$^{-1}$.

S20. (Movie): The ZZ(Pristine) related stretching and twisting vibration mode with frequency of 465.49 cm$^{-1}$.

S21. (Movie) **Mode-P2**: The ZZ(Pristine) edge related interlayer vertical vibrating with frequency of 352.59 cm$^{-1}$.

S22. (Movie) **Mode-U1**: The ZZ(U) related stretching vibration mode with frequency of 462.89 cm$^{-1}$.

S23. (Movie) **Mode-U2**: The breath mode in ZZ(U) with vibration frequency of 360.62 cm$^{-1}$.

# S1. Height of ZZ(Tube) and edge energies of five possible edge reconstructions for bilayer phosphorene

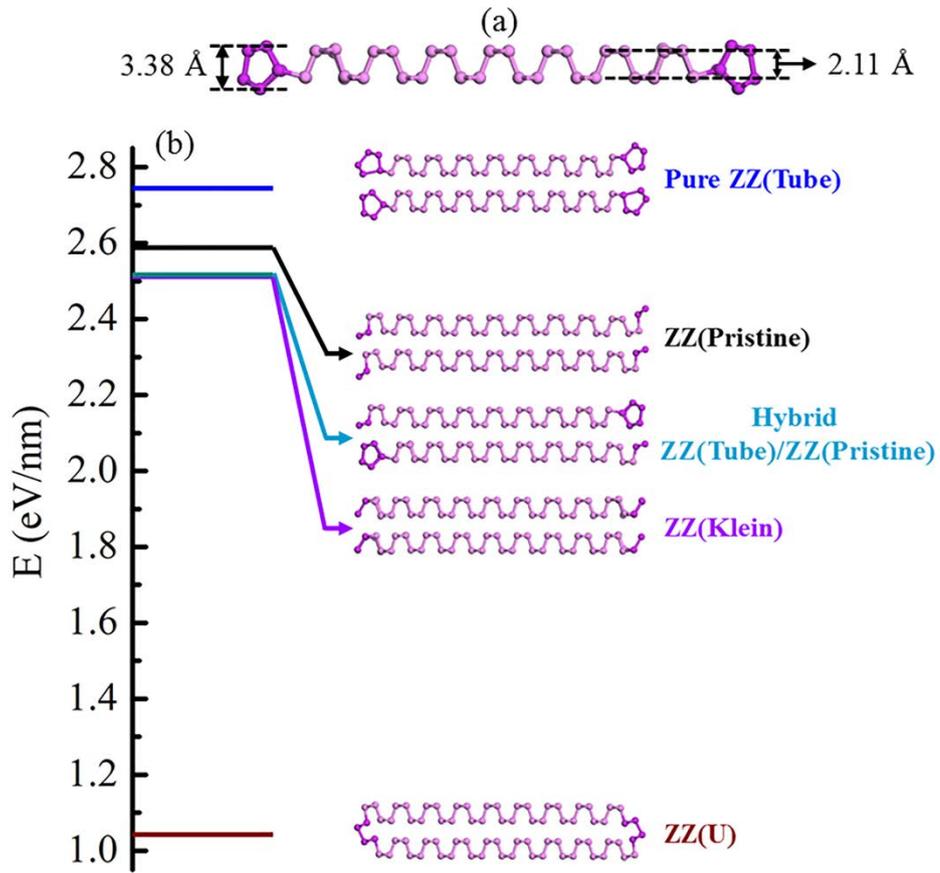

Figure S1 (a) The structure of monolayer PNR with ZZ(Tube). (b) The edge energies of Pure ZZ(Tube), ZZ(Pristine), Hybrid ZZ(Tube)/ZZ(Pristine), ZZ(Klein), and ZZ(U).

# S2. Evolution of temperature and total energy in the *ab initio* molecular dynamics (AIMD) simulations for ZZ(U)

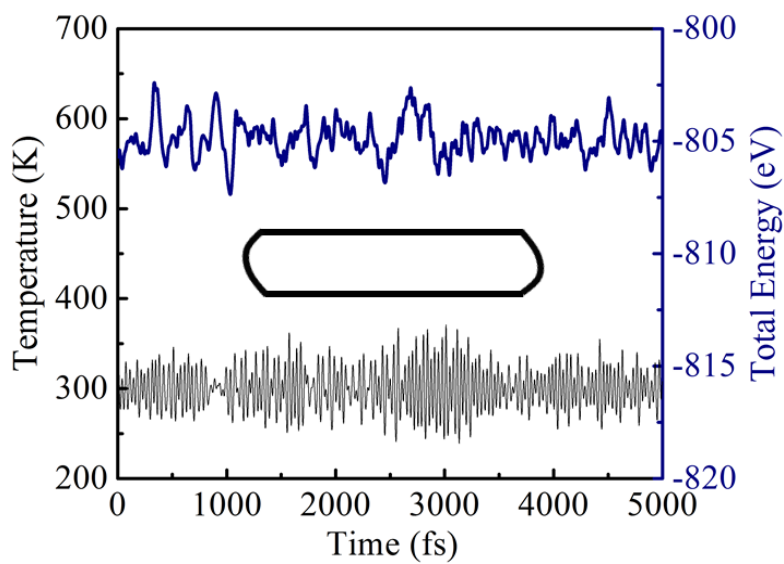

Figure S2 Temperature (black line) and total energy (blue line) profiles as a function of time during AIMD simulations for bilayer ZZ(U). The system is kept at 300 K for 5 ps.

# S3. Band structures of ZZ(Pristine) with width from 15.39 Å to 40.96 Å

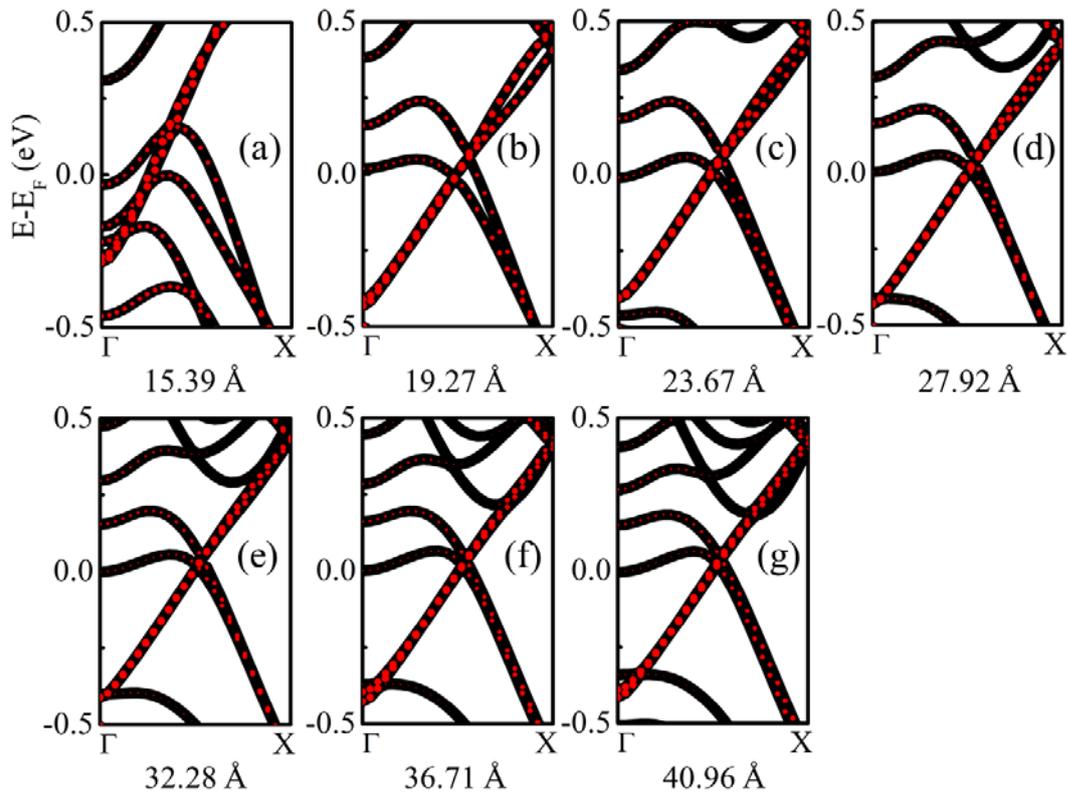

Figure S3 The electronic band structures with highlighted edge state as red dots for ZZ(Pristine) with width of 15.39 Å (a), 19.27 Å (b), 23.67 Å (c), 27.92 Å (d), 32.28 Å (e), 36.71 Å (f), and 40.96 Å (g), respectively.

# S4. Band structures of ZZ(Klein) with width from 14.67 Å to 40.52 Å

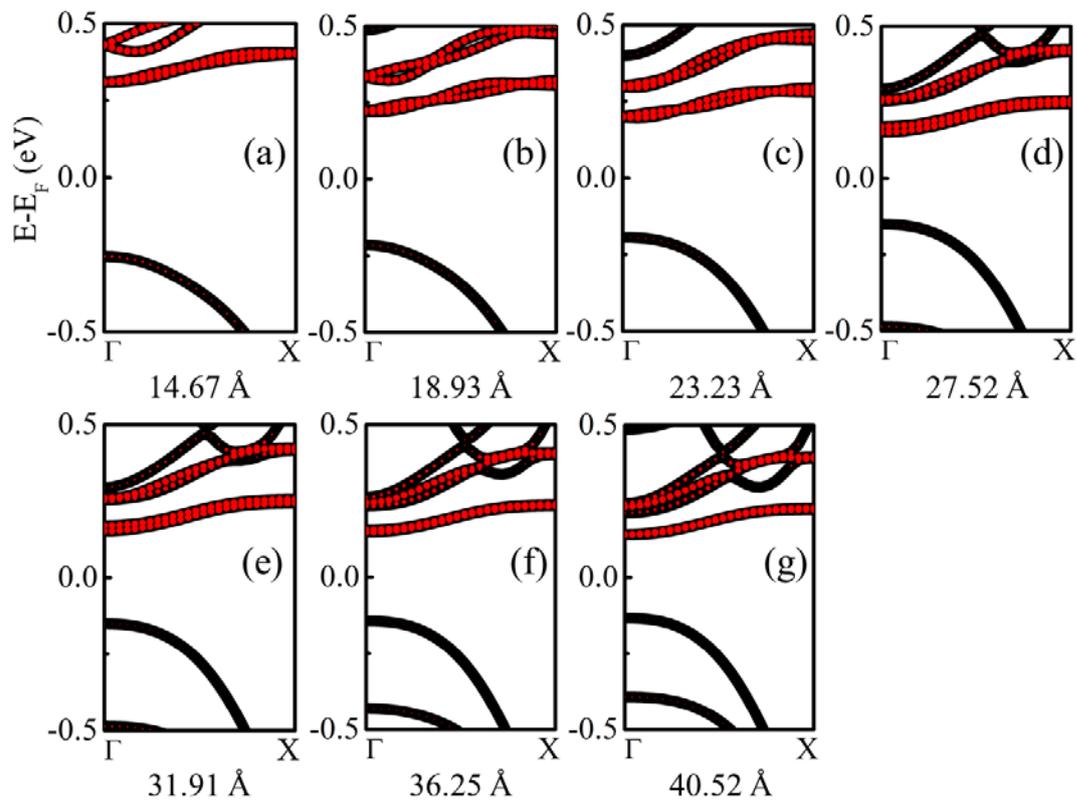

Figure S4 The electronic band structures with highlighted edge state as red dots for ZZ(Klein) with width of 14.67 Å (a), 18.93 Å (b), 23.23 Å (c), 27.52 Å (d), 31.91 Å (e), 36.25 Å (f), and 40.52 Å (g), respectively.

## S5. Band structures of ZZ(U) with width from 14.58 Å to 40.64 Å

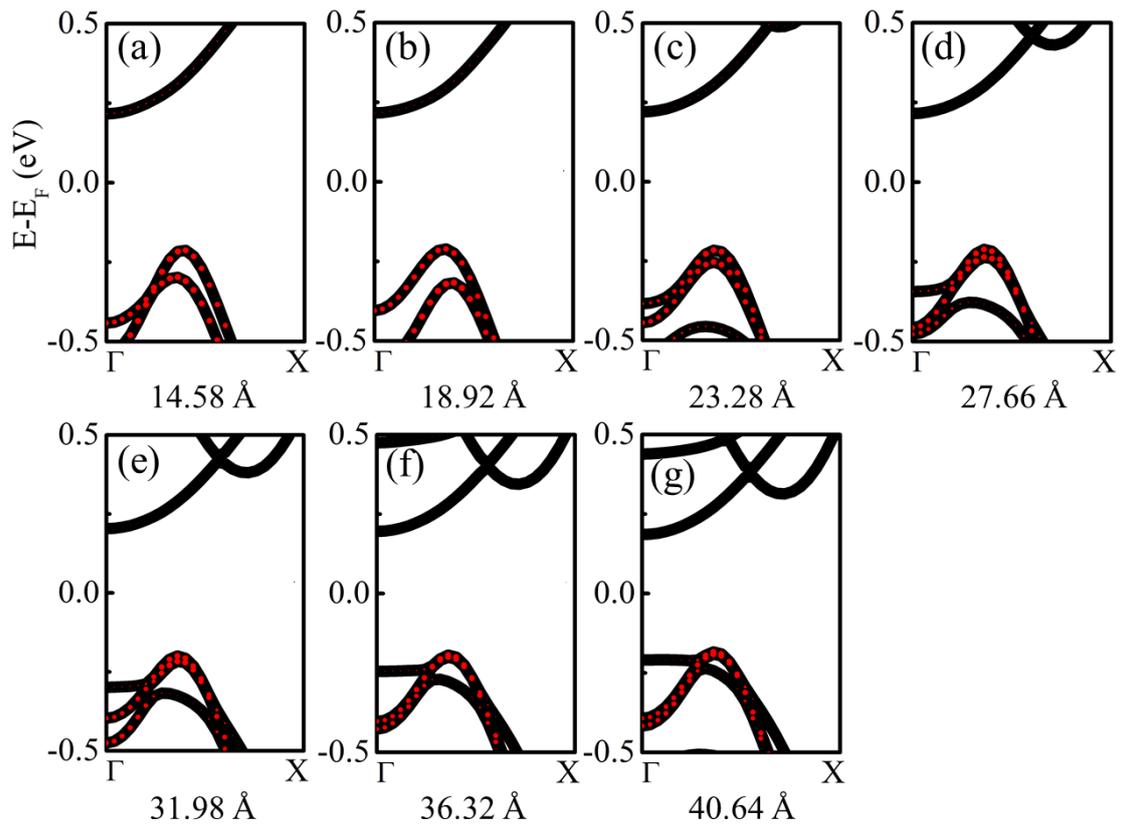

Figure S5 The electronic band structures with highlighted edge state as red dots for ZZ(U) with width of 14.58 Å (a), 18.92 Å (b), 23.28 Å (c), 27.66 Å (d), 31.98 Å (e), 36.32 Å (f), and 40.64 Å (g), respectively.

## S6. Band structure of the perfect 2L phosphorene

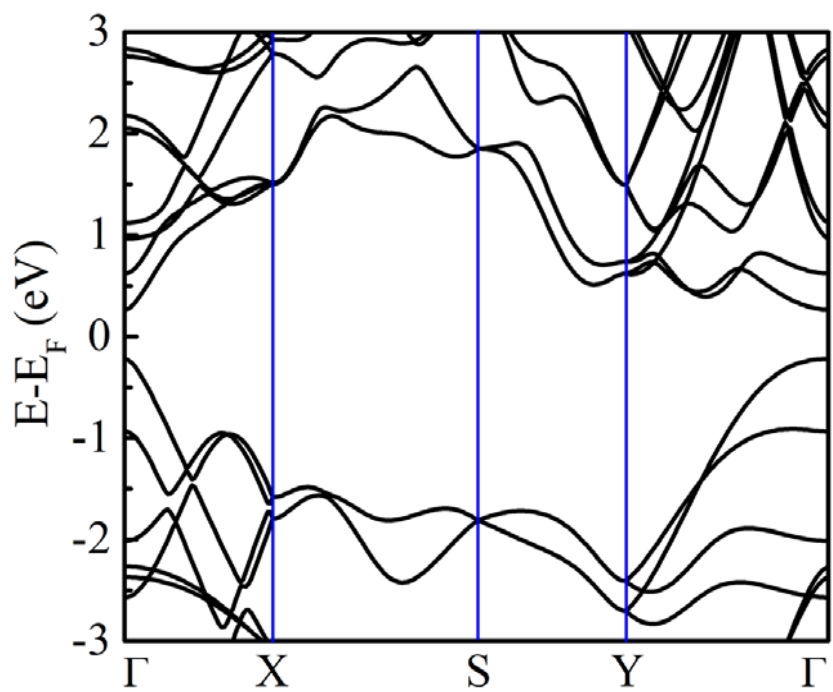

Figure S6 The electronic band structures of the perfect 2L phosphorene.

# S7. Band structures of hydrogenated ZZ(Pristine) with width from 17.27 Å to 43.15 Å

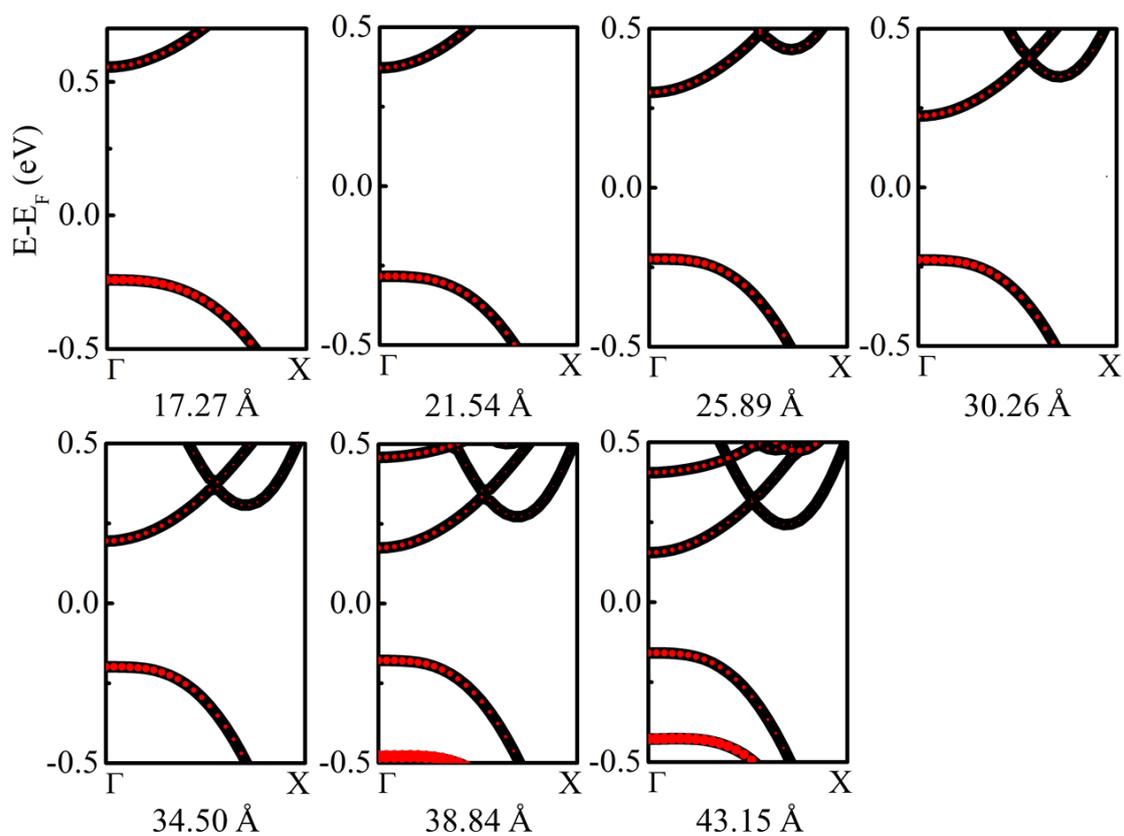

Figure S7 The electronic band structures with highlighted edge state as red dots for hydrogenated ZZ(Pristine) of width 17.27 Å (a), 21.54 Å (b), 25.89 Å (c), 30.26 Å (d), 34.50 Å (e), 38.84 Å (f), and 43.15 Å (g), respectively.

# S8. All the vibration frequencies for ZZ(Pristine) and ZZ(U) within (0, 518) cm$^{-1}$

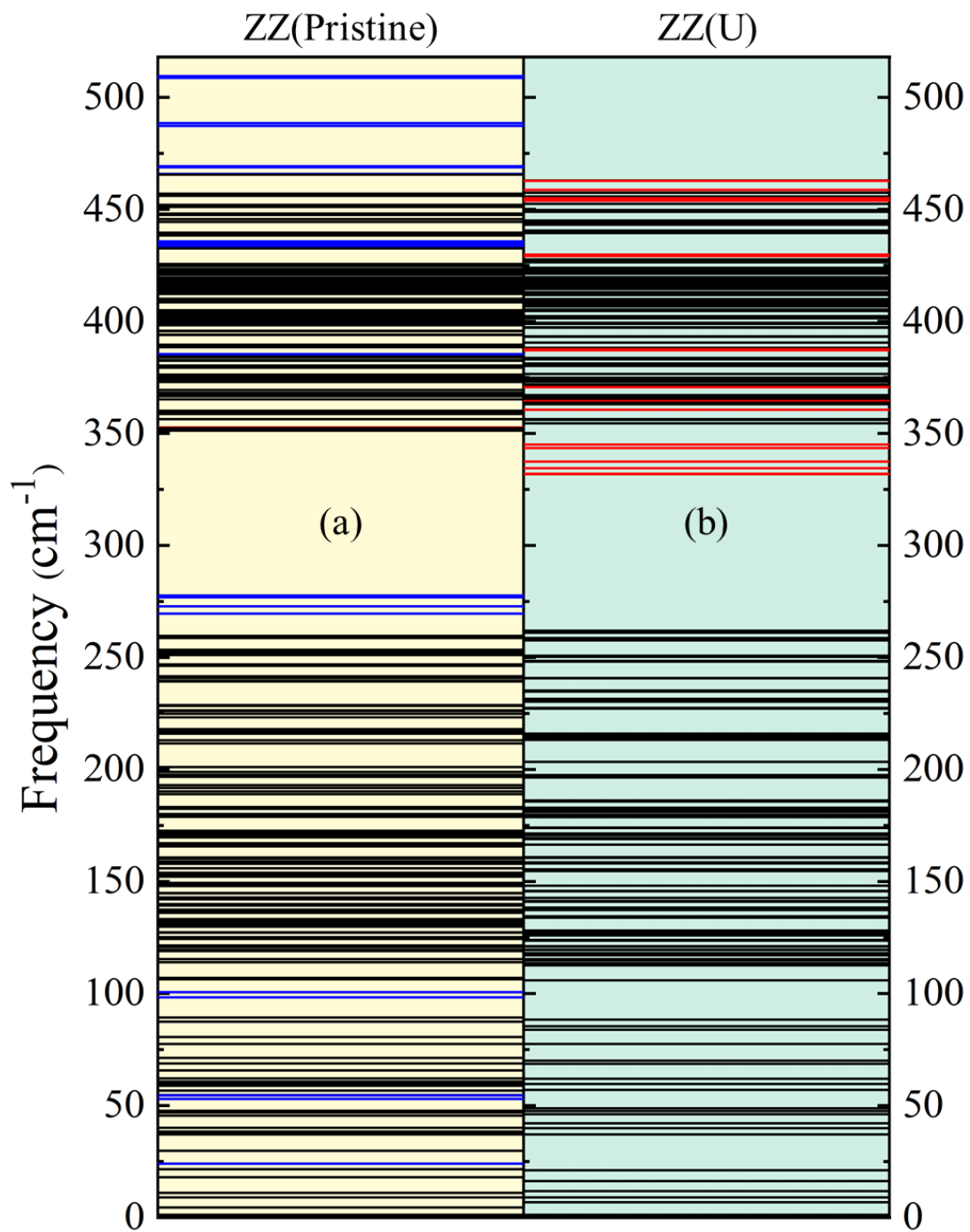

Figure S8 All the vibration frequencies for ZZ(Pristine) (a) and ZZ(U) (b) within (0, 518) cm$^{-1}$. The frequencies possessed by both ZZ(Pristine) and ZZ(U), only ZZ(Pristine), only ZZ(U) are denoted by black, blue and red lines, respectively.

## S9. Size effect on the thermal conductivity of PNRs

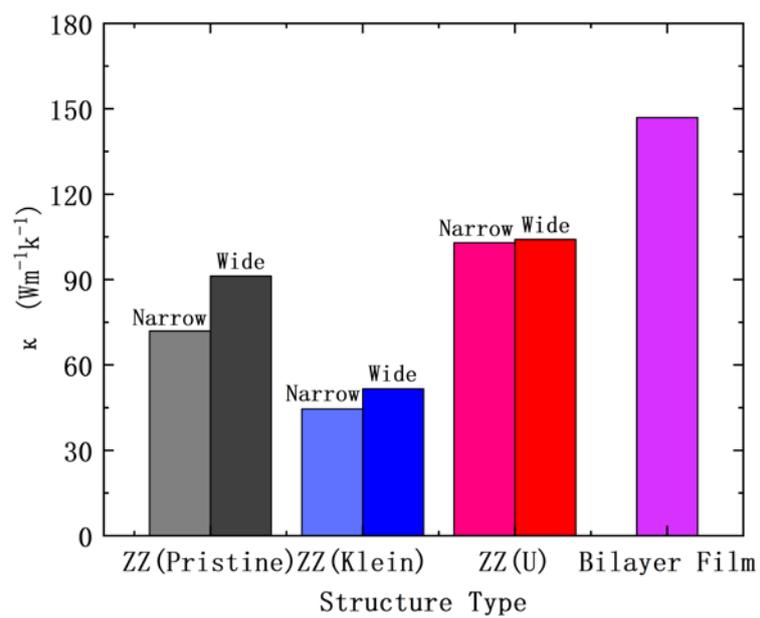

Figure S9 The thermal conductivities of PNRs with different width (narrow: 40.64 Å and wide: 62.28 Å).

# S10. Heat current autocorrelation function (HCACF) for pristine edge, Klein edge, and U-edge

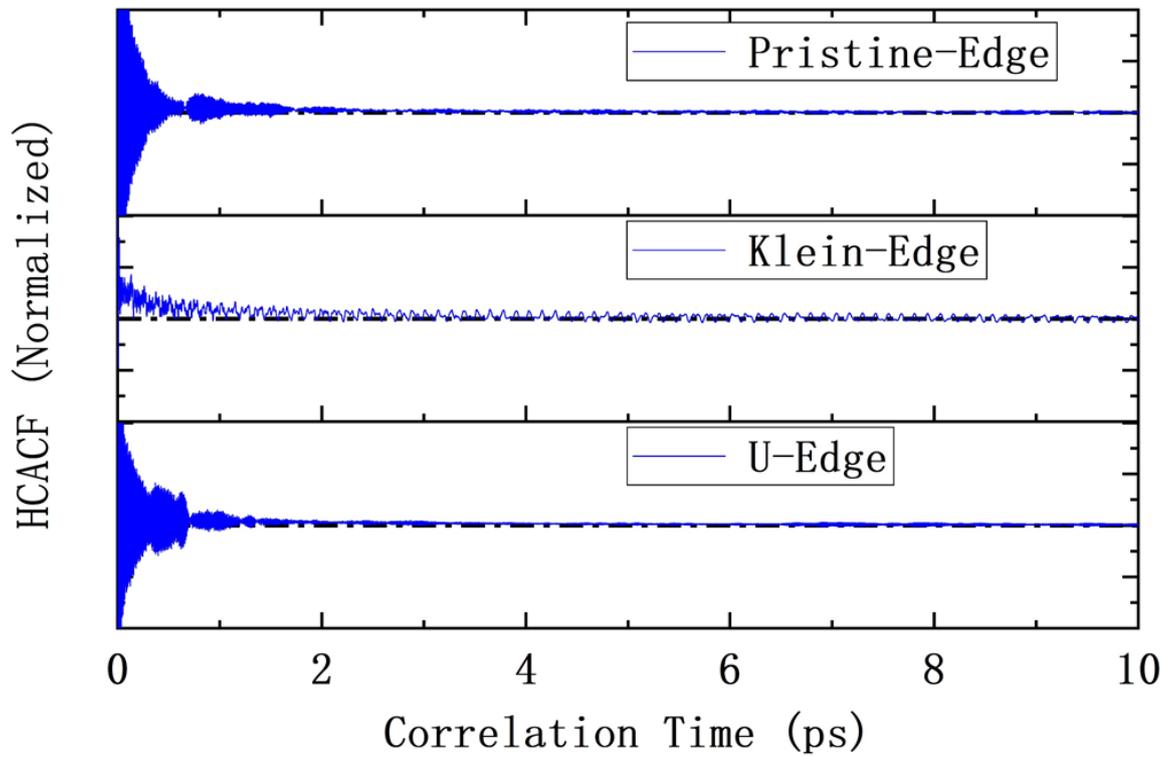

Figure S10 Normalized heat current autocorrelation function (HCACF) for pristine edge, Klein edge, and U-edge.

# S11. The thermal conductivity as a function of correlation time for pristine edge, Klein edge, and U-edge

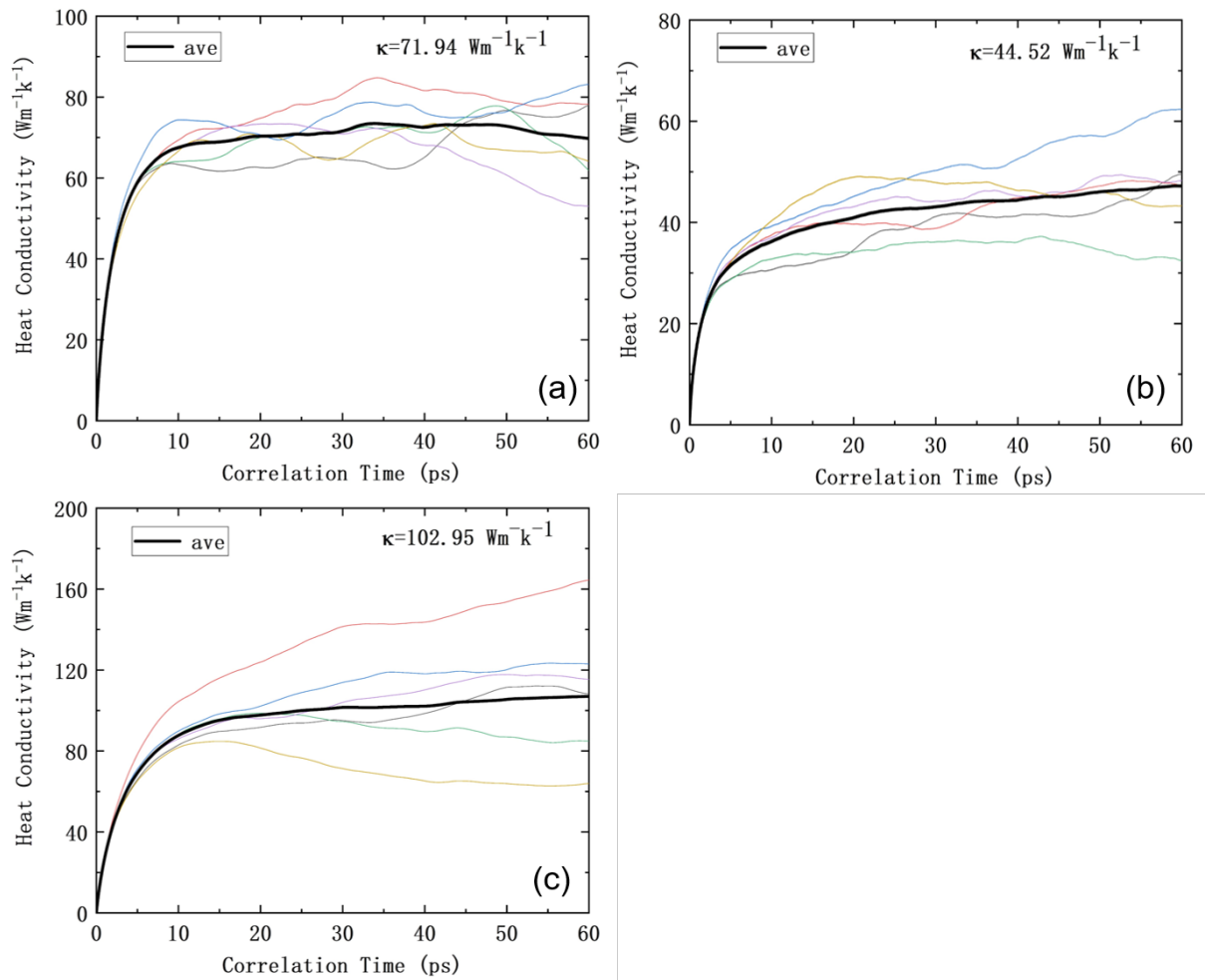

Figure S11 Thermal conductivity as a function of correlation time for pristine edge (a), Klein edge (b), and U-edge (c), respectively.